\documentclass{aa}
\usepackage{graphicx}
\usepackage[varg]{txfonts}
\usepackage{cases}
\usepackage{natbib}

\begin{document}

\title{Traveling kink oscillations of coronal loops launched by a solar flare}

\author{Dong~Li\inst{1,2}, Xianyong~Bai\inst{3,4}, Hui~Tian\inst{5,6}, Jiangtao~Su\inst{3,4}, Zhenyong~Hou\inst{5}, Yuanyong~Deng\inst{3,4}, Kaifan~Ji\inst{7}, and Zongjun~Ning\inst{1}}

\institute{Purple Mountain Observatory, Chinese Academy of Sciences, Nanjing 210023, PR China \email{lidong@pmo.ac.cn} \\
           \and Yunnan Key Laboratory of the Solar physics and Space Science,Kunming 650216, PR China \\
           \and National Astronomical Observatories, Chinese Academy of Sciences, Beijing 100101, PR China \email{xybai@bao.ac.cn} \\
           \and School of Astronomy and Space Sciences, University of Chinese Academy of Sciences, Beijing 100049, PR China \\
           \and School of Earth and Space Sciences, Peking University, Beijing, 100871, PR China \\
           \and Key Laboratory of Solar Activity and Space Weather, National Space Science Center, Chinese Academy of Sciences, Beijing 100190, PR China \\
           \and Yunnan Observatories, Chinese Academy of Sciences, Kunming 650011, PR China \\
           }
\date{Received; accepted}

\titlerunning{Kink oscillations of coronal loops}
\authorrunning{Dong Li et al.}

\abstract {Kink oscillations, which are often associated with
magnetohydrodynamic waves, are usually identified as transverse
displacement oscillations of loop-like structures. However, the
traveling kink oscillation evolving to a standing wave has rarely
been reported.} {We investigate the traveling kink oscillation
triggered by a solar flare on 2022 September 29. The traveling kink
wave is then evolved to a standing kink oscillation of the coronal
loop.} {The observational data  mainly come from the Solar Upper
Transition Region Imager (SUTRI), Atmospheric Imaging Assembly
(AIA), and Spectrometer/Telescope for Imaging X-rays (STIX). In
order to accurately identify the diffuse coronal loops, we applied a
multi-Gaussian normalization (MGN) image processing technique to the
extreme ultraviolet (EUV) image sequences at SUTRI~465~{\AA},
AIA~171~{\AA}, and 193~{\AA}. A sine function within the decaying
term and linear trend is used to extract the oscillation periods and
amplitudes. With the aid of a differential emission measure
analysis, the coronal seismology is applied to diagnose key
parameters of the oscillating loop. At last, the wavelet transform
is used to seek for multiple harmonics of the kink wave.} {The
transverse oscillations with an apparent decay in amplitude and
nearly perpendicular to the oscillating loop are observed in the
passbands of SUTRI~465~{\AA}, AIA~171~{\AA}, and 193~{\AA}. The
decaying oscillation is launched by a solar flare erupted close to
one footpoint of coronal loops and then it propagates along several
loops. Next, the traveling kink wave is evolved to a standing kink
oscillation. The standing kink oscillation along one coronal loop
has a similar period of $\sim$6.3~minutes at multiple wavelengths,
and the decaying time is estimated at $\sim$9.6$-$10.6~minutes.
Finally, two dominant periods of 5.1~minutes and 2.0~minutes are
detected in another oscillating loop, suggesting the coexistence of
the fundamental and third harmonics.} {First, we report the
evolution of a traveling kink pulse to a standing kink wave along
coronal loops that has been induced by a solar flare. We also
detected a third-harmonic kink wave in an oscillating loop. }

\keywords{Sun: flares ---Sun: oscillations --- Sun: coronal loop
--- Sun: UV radiation --- magnetohydrodynamics (MHD)}

\maketitle

\section{Introduction}
The solar corona, which lies in the upper atmosphere of the Sun, is
filled with various hot and magnetic structures, such as coronal
loops. These loop systems often reveal transverse oscillations, and
they are commonly connected to magnetohydrodynamic (MHD) waves in
the solar corona \citep[see][for a recent review]{Nakariakov20}. The
kink-mode oscillation, which is always perpendicular to the
oscillating loop and non-axisymmetric, is one of the most studied
MHD waves in the solar corona \citep{Nakariakov21,Lib23}. It was
first identified as the transverse displacement oscillation of
coronal loops in extreme ultraviolet (EUV) image sequences. Those
observed kink-mode oscillations were characterized by large-scale
amplitudes ($\gg$1~Mm) and quickly decaying within a few wave
periods, termed as "decaying oscillations"
\citep{Nakariakov99,Aschwanden02,Goddard16,Li17}. Later on, the
kink-mode oscillation without significant decaying was observed as
the transverse displacement in EUV images \citep{Wang12} or the
Doppler shift oscillation in coronal spectral lines \citep{Tian12}.
Such decayless oscillations often show small-scale amplitudes
($<$1~Mm) and can last for several wave periods or even many more
\citep{Anfinogentov15,Karampelas21,Mandal22}. Over the various
observations, kink-mode oscillations could be seen in nearly all the
loop-like structures, such as coronal loops, hot flare loops,
prominence threads, and even coronal bright points, since these
structures are all magnetic in nature; for instance, they all could
be regarded as thin magnetic flux tubes
\citep[e.g.,][]{Nakariakov99,Goossens13,Goddard16,Li18a,Li22a,Li23b,Nakariakov22,Zhang22}.
More interestingly, kink oscillations of a plasma slab could be seen
in microwave emissions \citep{Li20}, which  could be used to
explain the quasi-periodic pulsation at the wavelength of microwave
that is observed in the solar or stellar flare
\citep[e.g.,][]{Kaltman23}.

The kink-mode oscillation, in particular for the decaying
oscillation, is presumed to be excited by an impulsive solar
eruption, that is, a solar flare, a coronal jet, a flux rope, and so
on \citep[e.g.,][]{Zimovets15,Shen17,Shen18,Reeves20,Zhang20}. The
observed oscillation periods range from several minutes to a few
tens of minutes in duration, while the decaying time is roughly
equal to several oscillation periods
\citep{Goddard16,Nechaeva19,Ning22}. Conversely, decayless kink
oscillations have been demonstrated to be omnipresent in the solar
corona, but they appear to have no obvious connection to any solar
eruptive events
\citep[e.g.,][]{Tian12,Anfinogentov15,Nakariakov16,Guo22}. Their
displacement amplitudes are smaller than the minor radius of
oscillating loops, and their oscillation periods could range from a
few tens seconds to several hundreds seconds
\citep{Pascoe16,Li18b,Mandal21,Shi22,Zhong22}. For those standing
kink oscillations, their periods are strongly dependent on the loop
lengths, namely, a linear-growing relationship
\citep{Anfinogentov15,Guo20,Li23}. On the other hand, multiple
harmonics of standing kink oscillations were also observed in the
solar corona, in particular for the detection of the fundamental and
second harmonics
\citep[e.g.,][]{Verwichte04,McEwan08,Pascoe16,Duckenfield18}. In the
quiet-Sun loop, \cite{Duckenfield18} detected double periods of
$\sim$10.3~minutes and $\sim$7.4~minutes in the decayless
oscillation and they regarded them as the fundamental and second
harmonics of the standing kink wave. In another coronal loop, two
periods at $\sim$8~minutes and $\sim$2.6~minutes were simultaneously
seen in the decaying oscillation, which were explained as the
fundamental and third harmonics of the standing kink wave
\citep[e.g.,][]{Duckenfield19}. The detected period ratio of
multiple harmonics was always departure from unity, implying the
existence of density stratification along the oscillating loop
\citep{Andries05,Guo15}.

Kink oscillations have been well studied \citep[see][for a recent
review]{Nakariakov21}. However, the traveling kink wave evolved to a
standing kink oscillation is rarely observed. In this paper, we
explore an initial kink pulse launched by a solar flare and
propagating along several loops. The traveling kink wave is then
evolved to a standing kink oscillation within the fundamental and
third harmonics.

\section{Observations}
In this study, we mainly analyzed the EUV images taken by the Solar
Upper Transition Region Imager \citep[SUTRI;][]{Bai23} and the
Atmospheric Imaging Assembly \citep[AIA;][]{Lemen12} for the Solar
Dynamics Observatory (SDO). We also used X-ray fluxes recorded by
the Geostationary Operational Environmental Satellite (GOES) and the
Spectrometer/Telescope for Imaging X-rays \citep[STIX;][]{Krucker20}
on board the Solar Orbiter (SolO).

\begin{figure}[ht]
\centering
\includegraphics[width=\linewidth,clip=]{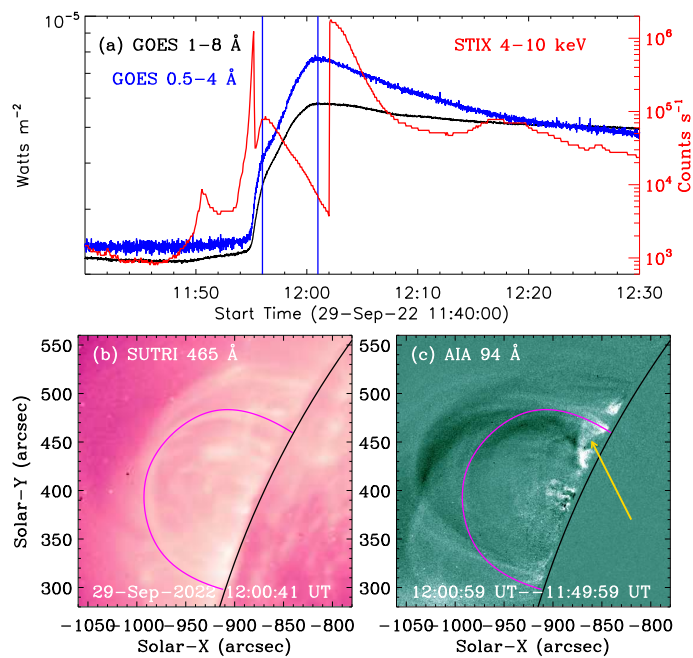}
\caption{Overview of the solar flare and coronal loop on 2022
September 29. (a): Light curves integrated over the entire Sun at
GOES 1$-$8~{\AA} (black) and 0.5$-$4~{\AA} (blue), and
STIX~4$-$10~keV (red). The blue vertical lines mark two peaks of the
solar flare in the GOES~0.5$-$4~{\AA} flux. (b) \& (c): Snapshots
with a FOV of $\sim$330\arcsec$\times$330\arcsec in passbands of
SUTRI~465~{\AA} and AIA~94~{\AA}. The magenta curve outlines a full
loop profile, the gold arrow indicates the flare site, and the
black line marks the solar limb. The whole evolution is shown in a
movie of anim.mp4. \label{over}}
\end{figure}

Figure~\ref{over} presents the overview of targeted coronal loops
and the associated flare on 2022 September 29. Panel~(a) shows GOES
fluxes at 1$-$8~{\AA} (black) and 0.5$-$4~{\AA} (blue), which
indicates a C5.7 class flare, it started at 11:50~UT and peaked at
12:01~UT. Interestingly, the GOES fluxes, particularly the SXR flux
at 0.5$-$4~{\AA} seems to have two main peaks (two vertical blue
lines), suggesting two episodes of energy releases. On the other
hand, the SXR light curve at 4$-$10~keV recorded by STIX suggests an
M4 class after considering the inserted attenuator
flare\footnote{https://datacenter.stix.i4ds.net/view/plot/lightcurves},
as shown by the red line in Figure~\ref{over}~(a). This is because
that STIX looked at the Sun from a different perspective than the
Earth, for instance, the angle between the Sun-SolO and Sun-Earth is
about 178.6$^{\circ}$. Thus, the solar flare considered here is
indeed a major flare.

The coronal loops were simultaneously observed by SUTRI and SDO/AIA
at wavelengths of EUV. SUTRI provides full-disk solar images at
Ne~VII~465~{\AA} with a formation temperature of about 0.5~MK
\citep{Tian17}, the pixel scale is $\sim$1.23\arcsec, and the time
cadence is roughly 30~s. Figure~\ref{over}~(b) shows the EUV image
taken by SUTRI at 12:00:41~UT, which shows several diffuse loops at
the solar north-east limb, and the magenta line outlines one entire
loop profile. Here, SUTRI successively observed the Sun from about
11:52~UT to 12:48~UT. SDO/AIA provides full-disk solar images at
seven EUV wavelengths with a time cadence of 12~s and each pixel has
a scale of 0.6\arcsec. Figure~\ref{over}~(c) shows the base
difference map (12:00:59~UT$-$11:49:59~UT) at AIA~94~{\AA}, which
shows bright emissions at one footpoint of the coronal loop, as
indicated by the gold arrow. The bright emissions can be regarded as
the major flare, which occurred at solar north-east limb, namely,
N26E86. However, it is hard to see any signatures of coronal loops
in the passband of AIA~94~{\AA}, largely because it contains
high-temperature plasma of $\sim$6.3~MK.

\section{Data reductions and results}
\subsection{Overview of coronal loops}
\begin{figure}[ht]
\centering
\includegraphics[width=\linewidth,clip=]{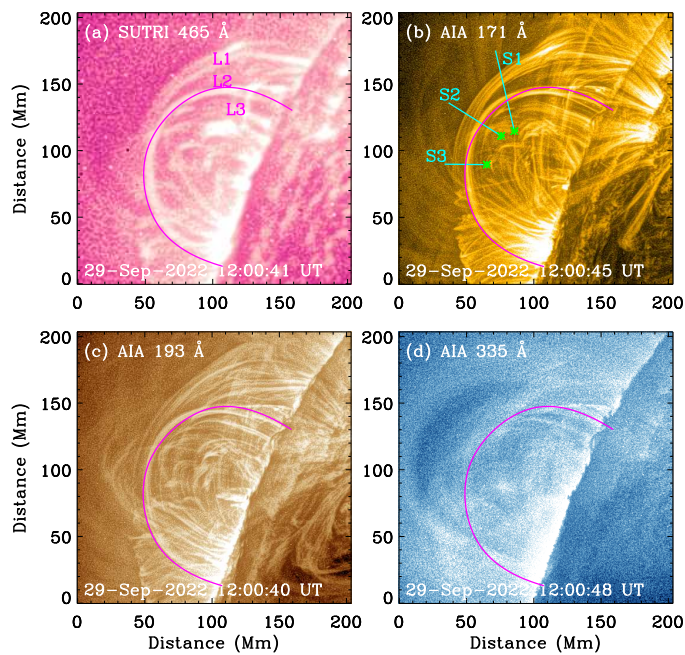}
\caption{Multi-wavelength images with a same FOV seen in
Figure~\ref{over}~(b). They are observed by SUTRI at 465~{\AA} (a),
SDO/AIA at 171~{\AA} (b), 193~{\AA} (c), and 335~{\AA} (d), and have
been processed by the MGN technique. The magenta curve outlines the
targeted loop. The straight cyan lines indicate the locations of
three artificial slits (S1, S2, and S3), which are used to generate
time-distance maps, and the green asterisks (`$\ast$') mark their
starting points. \label{imag}}
\end{figure}

In Figure~\ref{over}~(b), the coronal loops seen in the SUTRI map
appear to be very fuzzy, mainly due to the diffuse nature of EUV
emissions. In order to clearly identify these loop-like structures,
an image-processing technique such as a multi-Gaussian normalization
\citep[MGN;][]{Morgan14} was applied to the EUV image data observed
by SUTRI and SDO/AIA. Thus, the coronal loops are evidently
highlighted, as shown in Figure~\ref{imag}. A series of coronal
loops can be simultaneously seen in passbands of SUTRI~465~{\AA},
AIA~171~{\AA}, and 193~{\AA}. Herein, three coronal loops
indicated by L1, L2, and L3 are chosen to investigate the transverse
oscillation, since one of their footpoints is rooted in the flare
region. The studied coronal loops seem to consist of several
blended loops at AIA~171~{\AA} and 193~{\AA}, but those blended
structures can not be distinguished at SUTRI~465~{\AA}. Therefore,
we regard these coronal loops as loop systems and do not consider
their details such as fine-scale structures. On the other
hand, only one coronal loop reveals a complete loop profile, that
is, the loop apex and double footpoints can be clearly seen in
EUV maps, which is regarded as the targeted loop (L2), as outlined
by the magenta curve. While the other two loops (i.e., L1 and L3)
just show one footpoint and the loop apex. Similarly to what has
observed at AIA~94~{\AA}, those coronal loops can not be well seen
at AIA~335~{\AA}, as shown in panel~(d). Our observations
suggest that the loop systems only contain plasma at low
temperatures, that is, $<$2~MK.

The movie anim.mp4 shows the whole evolution of coronal loops and
the associated flare from $\sim$11:52~UT to $\sim$12:18~UT. From
this, we can find that at about 11:56~UT, a solar flare erupts close
to the northern footpoint of the loop systems (see Fig.~\ref{over})
and then it subsequently induces a transverse oscillation in the
targeted loop system. Interestingly, the transverse oscillation
appears to propagate along several loops, namely, from L3 through L2
to L1, and then it evolves to a standing kink oscillation. The
transverse oscillation continues to exist until around 12:10~UT,
when the coronal loops gradually disappear. In order to take a
closer look the appearance of the transverse oscillation, we
generated time-distance (TD) maps along three artificial straight
slits (S1, S2, and S3), which are nearly perpendicular to the loop
axis. This is indicated by three cyan lines in Figure~\ref{imag} and
the green stars (`$\ast$') mark their starting points.

\subsection{Time-distance maps}
\begin{figure}[ht]
\centering
\includegraphics[width=\linewidth,clip=]{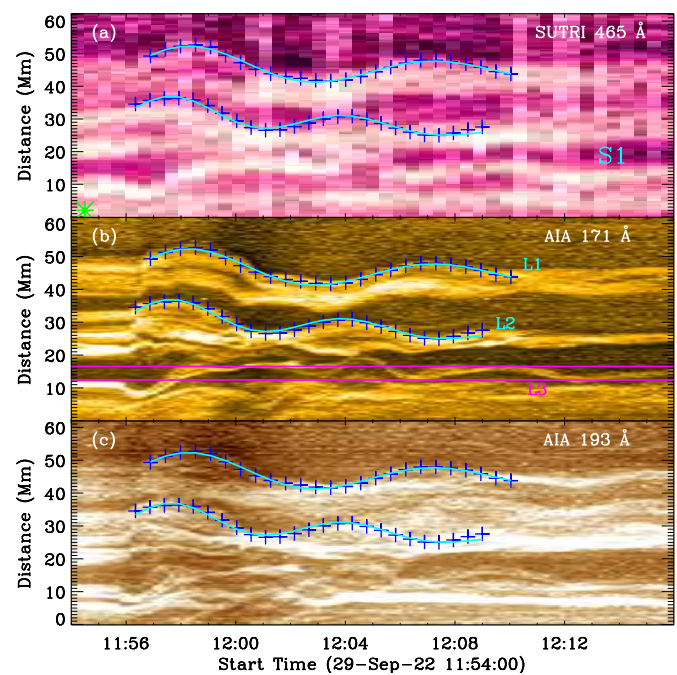}
\caption{TD maps along slit~S1 from 11:54~UT to 12:16~UT, which are
made from image sequences at SUTRI~465~{\AA} (a), AIA~171~{\AA} (b),
and 193~{\AA} (c). These blue pluses (`$+$') highlight the skeletons
of oscillating structures, whereas the cyan curves represent their
best fitting results. The green asterisk (`$\ast$') marks the starting
point of the TD map. Two magenta lines label another oscillating
loop seen at AIA~171~{\AA}. \label{slit1}}
\end{figure}

Figure~\ref{slit1} presents TD maps for slit S1 that crosses the
coronal loops in passbands of SUTRI~465~{\AA}, AIA~171~{\AA} and
193~{\AA}, and the green symbol of `$\ast$' indicates the zero-point
of $y$-axis. In order to avoid any confusions, the slit S1 is
selected at the northern locations where there are less overlaps
with neighboring loops, and it is close to the solar flare. In these
multi-wavelength TD maps, one can immediately notice that several
transverse oscillations within at least two peaks. We first analyze
the TD map at SUTRI~465~{\AA}, because it shows two apparent
transverse oscillations in loops L1 and L2 from about 11:56~UT to
12:10~UT. The oscillating locations of coronal loops are often
determined by a Gaussian fitting method
\citep[e.g.,][]{Wang12,Zhong22}. However, it is impossible to use
this method if several overlapping loops simultaneously appear in
the TD map \citep[cf.][]{Anfinogentov15,Goddard16}. Therefore, we
manually identified the edge of the oscillating loop
\citep[cf.][]{Gao22} along the transverse direction as the
oscillatory locations, as marked by the blue pluses (`+') in
panel~(a). The two transverse oscillations appear to decay weakly,
so a combination of a sine function, a decaying term and a linear
trend is used to fit the loop oscillation
\citep[e.g.,][]{Nakariakov99,Goddard16,Su18}, as shown
by~Equation~\ref{eq1}:

\begin{equation}
  y(t)=A \cdot \sin(\frac{2 \pi}{P}~t+ \psi ) \cdot e^{-\frac{t}{\tau}}+ k \cdot t + C,
\label{eq1}
\end{equation}

\noindent Here, $A$ represents the initial displacement amplitude,
$P$ and $\tau$ stand for the oscillation period and decaying time,
$\psi$ and $C$ are initial phase and location of the transverse
oscillation, and $k$ is a constant that refers to the drifting
velocity of the oscillating loop system in the plane-of-sky. The
fitting results are indicated by the cyan curve in
Figure~\ref{slit1}~(a), which match well with those identified
skeletons of the oscillating loop system. Next, we could determine
the velocity amplitude ($v_m$) by using the derivative of the
displacement amplitude \citep[cf.][]{Gao22,Li22a}, such as $v_m=2\pi
\cdot \frac{A}{P}$. Some key parameters measured in the transverse
oscillation are listed in Table~\ref{loop}.

Figure~\ref{slit1}~(b and c) presents TD maps in passbands of
AIA~171~{\AA} and 193~{\AA}, respectively. Besides the two
transverse oscillations seen in the passband of SUTRI~465~{\AA}, we
can also see some other transverse oscillations, one such case is
outlined by two magenta lines, that is, L3. However, the
displacement profile is very different from a sine function, which
could be considered as the signature of multiple harmonics and set
it aside analysis later. Herein, we first focus our attention on the
apparent transverse oscillations (L1 and L2), and these selected
oscillating locations in panel~a are directly overplotted in TD maps
at AIA~171~{\AA} and 193~{\AA}, as shown by blue pluses in panels~b
and c. They appear to match well with the profile of transverse
oscillations, suggesting a multi-thermal nature of the oscillating
loop system. The best fitting results indicated by the cyan curves
confirm that the transverse oscillation is basically a manifestation
of decaying kink oscillation.

\begin{figure}[ht]
\centering
\includegraphics[width=\linewidth,clip=]{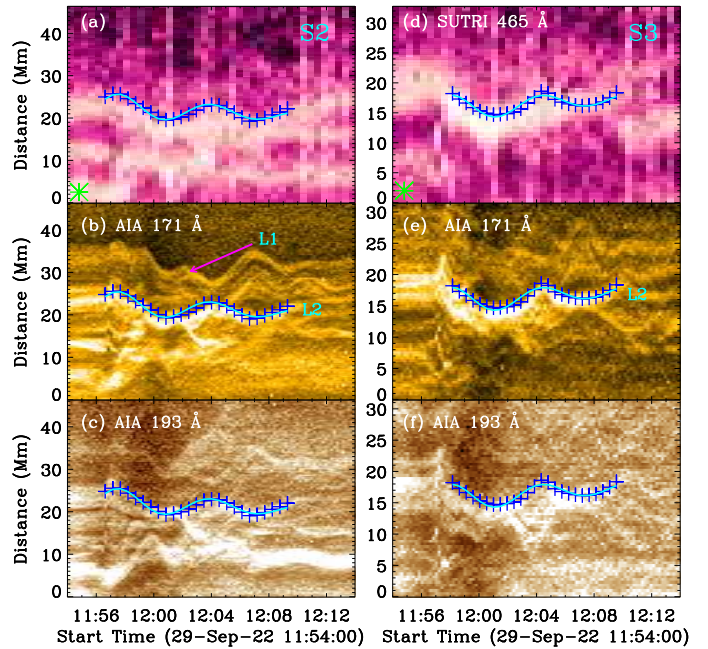}
\caption{Similar to Figure~\ref{slit1}, but they are generated from
slits~S2 (a$-$c) and S3 (d$-$f). The magenta arrow mark the
transverse oscillation with multiple harmonics.  \label{slit2}}
\end{figure}

Figure~\ref{slit2} shows TD maps along two straight slits that
cross the loop apex (S2) and the southern loop leg (S3), and they
are almost perpendicular to the loop axis. Similarly to slit S1, an
apparent transverse oscillation of loop L2 is simultaneously
observed in passbands of SUTRI~465~{\AA}, AIA~171~{\AA,} and
193~{\AA}. The oscillating locations are manually selected from the
TD map at SUTRI~465~{\AA}, and they could also appear in TD maps at
AIA~171~{\AA} and 193~{\AA}, as indicated by blue pluses. The
observations suggest that the transverse oscillation can be
observed in multi-thermal loop system and no apparent phase
difference appears at these three passbands. Table~\ref{loop} also
lists some key parameters for the transverse oscillation. Almost the
same oscillation period implies that the transverse oscillation seen
in three straight slits comes from the same oscillating loop system.
We also notice that the transverse oscillation of loop L1
can only be seen at AIA~171~{\AA} at the loop apex (S2), and it
almost disappears at the southern loop leg (S3). Moreover, the
displacement profile is also different from the sine function,
implying a multi-harmonics wave, which is similar to the transverse
oscillation of loop L3. The transverse oscillation of loop L3 can
not be seen in slits S2 and S3, mainly because that the coronal loop
L3 disappears. Thus, only the loop L2 that has an entire profile is
detailed analyzed, as shown in table~\ref{loop}.

\begin{table}
\caption{Key parameters measured in the oscillating loop L2 at three positions.}
\label{loop}      
\tabcolsep 10pt        
\begin{tabular}{ccccccc}
\hline
                        &  S1     & S2       &  S3       \\
\hline
$L$ (Mm)                &  221.3  &  221.3   &  221.3    \\
$P$ (minutes)           &  6.32    &  6.38     &  6.32       \\
$\tau$ (minutes)        &  9.6    &  10.5    &  10.6      \\
$A$ (Mm)                &  12.5   &  7.9     &  6.1       \\
$v_{\rm m}$ (km~s$^{-1}$)  &  207 &  130     &  101     \\
$c_{\rm k}$ (km~s$^{-1}$)  &  1167  &  1156  &  1167      \\
$n_{\rm i}$ (cm$^{-3}$)    &  1.93$\times$10$^9$  & 1.66$\times$10$^9$  & 1.61$\times$10$^9$     \\
$n_{\rm e}/n_{\rm i}$      &  0.30   &  0.35    &  0.36     \\
$v_{\rm A}$ (km~s$^{-1}$)  &  941    &  950     &  962      \\
$B$ (G)                 &  21.3   &  20.0    &   19.9    \\
\hline
\end{tabular}
\end{table}

\begin{figure}[ht] \centering
\includegraphics[width=\linewidth,clip=]{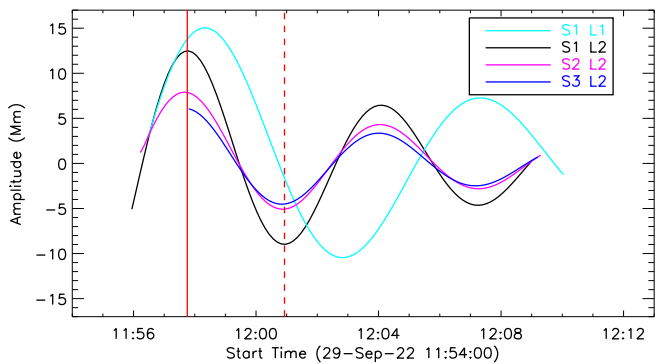}
\caption{Best-fitting functions after removing the linear trend
for the artificial slits~S1 (black), S2 (magenta), and S3 (cyan) in
coronal loops L1 and L2. \label{corr}}
\end{figure}

In Figure~\ref{corr}, we show the best fitting results from the
transverse oscillations that are generated from three straight slits
in coronal loops of L1 and L2. The linear trend has been removed, so
that they can be directly compared in the same window. Along
the same slit S1, a visible time difference is seen when the
transverse oscillation goes through loop L2 (black) and L1 (cyan),
implying that the transverse oscillation is propagating along these
two coronal loops. We can also find that the oscillation period in
loop L1 is obviously longer than that in loop L2, because that the
loop L1 is much longer than L2, as seen in Figure~\ref{imag}. In
the same loop L2, the transverse oscillation at three different
positions reaches the maximum (red solid line) and minimum (red
dashed line) at almost the same time, suggesting that the loop
system oscillates nearly in-phase along the loop length. The
displacement amplitude of the transverse oscillation at the northern
loop leg (S1) is obviously larger than that at the loop apex (S2)
and at the southern loop leg (S3), because the solar flare that
triggers the transverse oscillation erupted near the northern
footpoints of the loop system, as shown in Figure~\ref{over}. Our
observations also suggest that the transverse oscillation of loop L2
is indeed the fundamental mode.

\begin{figure}[ht]
\centering
\includegraphics[width=\linewidth,clip=]{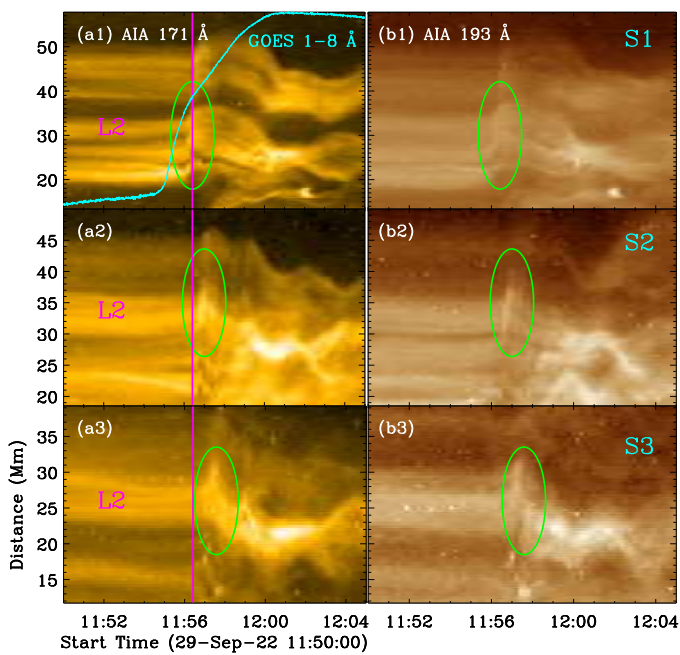}
\caption{TD maps along slits~S1$-$S3 from 11:50~UT to 12:05~UT,
generated from data series at wavelengths of AIA~171~{\AA}
(a1$-$a3) and 193~{\AA} (b1$-$b3). The magenta line marks a fixed
time, and the green ellipse outlines the initial short pulse in
different slits. The overplotted cyan line represents the
GOES~1$-$8~{\AA} flux. \label{slit3}}
\end{figure}

Figure~\ref{corr} demonstrates that the transverse oscillations in
different loops are out of phase. However, it does not illustrate
that the initial pulse triggered by the solar flare is a traveling
wave, which is well seen in the movie anim.mp4. In order to provide
the adequate demonstration, Figure~\ref{slit3} presents the TD plots
taken from different slits, which are generated from AIA~171~{\AA}
and 193~{\AA} image series during 11:50$-$12:05~UT. The overlaid
cyan line is the GOES SXR light curve at 1$-$8~{\AA}. We note that
these TD images have been zoomed and have a common time axis. The
coronal loop L2 is visible in these TD images at both AIA~171~{\AA}
and 193~{\AA}, and it does not show any signature of transverse
oscillations before the solar flare, for instance, from 11:50~UT to
11:55~UT. Then, a short transverse pulse appears in the coronal
loop, which is accompanied by the flare eruption, as indicated by
the green ellipse and cyan line. Next, the short pulse, which could
be regarded as an initial transverse pulse, is evolved to a standing
transverse oscillation of the coronal loop, as shown in
Figure~\ref{slit1}. Interestingly, the initial short pulse in the
loop L2  appears later and later from slits S1 to S3, suggesting
that there is a noticeable time delay between the appearance of the
initial transverse pulse in different slits, as indicated by the
green ellipse and magenta line in panels~(a1)$-$(a3). The similar
short transverse pulse with a time delay between different slits can
also be seen at AIA~193~{\AA}, as shown in panels~(b1)$-$(b3). All
those observations demonstrate that the initial short pulse is
traveling along the coronal loop. That is to say, the initial pulse
launched by the flare is a traveling wave.

\subsection{DEM results}
\begin{figure}[ht]
\centering
\includegraphics[width=\linewidth,clip=]{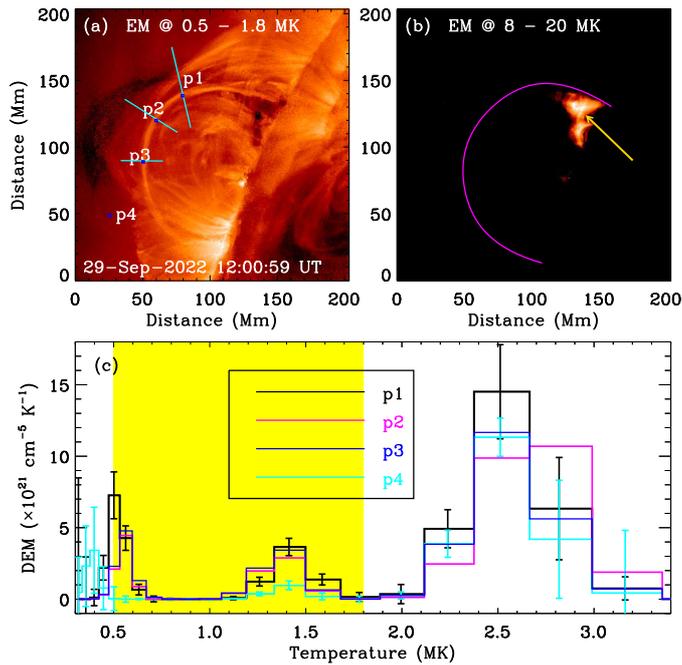}
\caption{DEM analysis of the oscillating loop. (a) \& (b):
narrow-band EM images integrated in temperature ranges of
0.5$-$1.8~MK and 8$-$20~MK. (c): DEM profiles at three locations
(P1, P2, and P3) inside the oscillating loop and at one location
(P4) away from the oscillating loop. The yellow region outlines the
EM and number density integration range for the oscillating loop.
The error bars denote to the uncertainties ($3\delta$) of the DEM
solution. \label{dem}}
\end{figure}

We further perform the differential emission measure (DEM) analysis
for oscillating loop systems and the associated flare, as shown in
Figure~\ref{dem}. In this study, an improved sparse-inversion code
\citep{Cheung15} developed by \cite{Suy18} was applied to determine
the DEM($T$) distribution at every pixel, which is calculated from
the SDO/AIA image data at six EUV passbands, that is, AIA~94~{\AA},
131~{\AA}, 171~{\AA}, 193~{\AA}, 211~{\AA}, and 335~{\AA}. Their
uncertainties are estimated from 100 Monte Carlo (MC) simulations
for each pixel, that is, 3$\delta$ ($\delta$ refers to the standard
deviation of 100~MC simulations). Panels~(a) and (b) show
narrow-band EM images that are integrated in temperature ranges of
0.5$-$1.8~MK and 8$-$20~MK, respectively. We immediately notice that
coronal loops can be clearly seen at the lower temperature range
between 0.5$-$1.8~MK (panel~a), while the solar flare near the
northern footpoint of the loop system is prominently visible at the
higher temperature range of 8$-$20~MK, as marked by the gold arrow
in panel~(b). We also notice that only loop L2 has the whole
loop-like profile in the EM map, which is consistent with SUTRI and
SDO/AIA observations.

Figure~\ref{dem}~(c) shows the DEM profiles with error bars such as
3$\delta$ as a function of temperature. Here, we choose three
positions (p1, p2, and p3) inside the oscillating loop (L2) and one
position (p4) that is away from the oscillating loop (or background
corona), as indicated by the blue boxes in panel~(a). For clarity,
only the error bars at the northern loop region (black line) and at
the background position (magenta line) are shown in panel~(c). The
DEM profiles inside the oscillating loop (p1, p2, and p3) exhibit
three apparent peaks at about 0.5~MK, 1.4~MK, and 2.5~MK, while the
DEM profile at background corona (p4) only has two prominent peaks
at around 0.4~MK and 2.5~MK. Moreover, the high-temperature peak at
about 2.5~MK are roughly equal at those four positions, suggesting
that it does indeed emit from the coronal emission of the diffuse
background. On the other hand, the low-temperature peak at roughly
0.4~MK from the coronal background is significantly away from the
peak at $\sim$0.5~MK. Based on these facts, we can conclude that the
oscillating loop system of interest covers a temperature range from
about 0.5~MK to 1.8~MK, as indicated by the yellow shadow. This
agrees with our imaging observations, for instance, the oscillating
loop system is clearly seen in passbands of SUTRI~465~{\AA}
($\sim$0.5~MK), AIA~171~{\AA} ($\sim$0.63~MK), and 193~{\AA}
($\sim$1.58~MK).

\subsection{Coronal seismology}
Generally, the transverse oscillation of an entire coronal loop is
regarded as kink-mode wave because that the global sausage-mode
wave requires for the very thick loop with quite denser plasmas
\citep{Nakariakov03,Tian16}. Herein, we performed an MHD coronal
seismology with Equations (2)$-$(4), based on the fundamental
kink-mode oscillation of the coronal loop L2
\citep[cf.][]{Van14,Yuan16,Long17,Yang20,Nakariakov21}.

\begin{eqnarray}
  c_{\rm k} &=& \frac{2L}{P}, \\
  v_{\rm A} &=& c_{\rm k} \cdot \sqrt{\frac{1+n{\rm_e}/n{\rm_i}}{2}}, \\
  B & \approx & v_{\rm A} \cdot \sqrt{\mu_0~n_{\rm i}~m_{\rm p}~\widetilde{\mu}},
  \label{eq2}
\end{eqnarray}

\noindent Here $c_{\rm k}$ is the kink speed, $L$ refers to the
length of oscillating loop, which can be determined by the distance
between double footpoints when assuming a semi-circular shape for
the coronal loop \citep[cf.][]{Tian16,Li22b}, as indicated by the
magenta curve in Figures~\ref{imag} and \ref{dem}. $n{\rm_e}$ and
$n{\rm_i}$ represent external and internal number densities of the
coronal loop, and they could be determined by the DEM results, such
as $\sqrt{EM/w}$. $w$ is the integration length, which could
consider as the full width at the half maximum of the coronal loop
along its cross section, while it is the effective line-of-sight
depth ($w \approx 4 \times 10^{10}$~cm) in the background corona
\citep{Zucca14,Su18}. $v_{\rm A}$ and $B$ are the local Alfv\'{e}n
speed and magnetic field strength inside the oscillating loop
system. Also, $\mu_0$ is the magnetic permittivity in vacuum, and
$\widetilde{\mu}$ ($\approx$1.27) stands for the effective particle
mass with respect to the proton mass ($m_{\rm p}$) in the solar
corona \citep[cf.][]{White12}. The estimated parameters parameters
are listed in Table~\ref{loop}.

\subsection{Multiple harmonics}
\begin{figure}[ht]
\centering
\includegraphics[width=\linewidth,clip=]{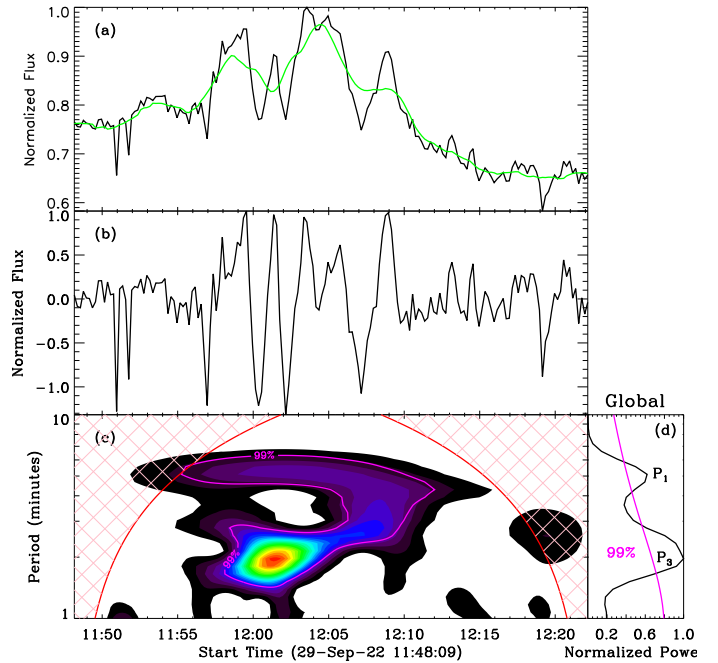}
\caption{Wavelet analysis results. (a): Normalized time series
integrated over two magenta lines in Figure~\ref{slit1}, the
overlaid green line is the slow-varying trend. (b) Normalized
detrended time series. (c): Morlet wavelet power spectrum. (d):
Global wavelet power spectrum. The magenta contours indicate a
significance level of 99\%. \label{wav}}
\end{figure}

In Figures~\ref{slit1}~(b) and \ref{slit2}~(b), we can find that the
displacement profile is very different from a sine, which is a
strong signature of multiple harmonics of kink waves. Therefore,
their periods are difficult to be determined by fitting a sine
function, such as Equation~1. In order to identify the multiple
periods, we performed a wavelet transform \citep{Torrence98} for the
time series of oscillating loop L3, as shown in Figure~\ref{wav}.
Panel~(a) presents the raw time series integrated over two magenta
lines in Figure~\ref{slit1}~(b), and the time series has been
normalized by its peak value. The overlaid green line represents the
slow-varying trend, and the detrended time series is shown in
panel~(b). Here, we used a smooth window of 3~minutes to obtain the
slow-varying trend (green line), because we thereby enhance the
short-period oscillation and suppress the long-period trend
\citep[e.g.,][]{Kupriyanova13,Kolotkov16,Li21,Lid22}. Panels~(c) and
(d) shows the Morlet wavelet power spectrum and global wavelet power
spectrum, respectively. From which, we can identify at least two
periods with large uncertainties. Then, two dominant periods of
5.1~minutes (P$_1$) and 2.0~minutes (P$_3$) are determined by the
double peaks above the 99\% significance level in the global wavelet
power spectrum. The period ratio (P$_1$/3P$_3$) is estimated to be
0.85, similarly to what has found between the fundamental and third
harmonics in the decaying kink oscillation
\citep[cf.][]{Duckenfield19}. So, the kink oscillation of loop L3
contains a third harmonic. Similarly, the standing kink oscillation
of loop L1 along slit S2 also shows a strong signature of multiple
harmonics, which is evolved from the traveling kink wave launched by
a solar flare, as shown in Figure~\ref{slit1}.

\section{Conclusion and discussion}
Using the EUV images taken by SUTRI and SDO/AIA, we investigate the
decaying transverse oscillation of coronal loops. Combined
observations from GOES and STIX reveal a major flare  erupted close
to one footpoint of those oscillating loops.

The observed oscillation of coronal loops is transverse in
nature. It lasts for at least two wave periods while significantly
decaying in amplitude; that is to say, the observed oscillation is
basically an decaying kink wave. The initial displacement amplitude
could be as large as 12.5~Mm and it decays rapidly, which is
similar to previous findings about the decaying oscillation of
coronal loops
\citep[e.g.,][]{Nakariakov99,Goddard16,Su18,Nechaeva19,Ning22},
confirming that the transverse oscillation is indeed a decaying kink
wave.

A solar flare is simultaneously observed near the northern footpoint
of the oscillating loops. It is a C5.7 flare according to the GOES
SXR classification, while it is an M4 class measured by the STIX
4$-$10 flux. This is because that the flare located at the solar
limb from the Earth-orbit perspective, so only a partial emission
could be received by GOES. STIX measured the whole flare emission at
X-ray band at a different perspective. For this reason, STIX light
curves are inserted attenuator. In any case, the solar flare is a
major flare and it induces the decaying kink oscillation of coronal
loops. This is also similar to pervious observations, for instance,
the decaying oscillation is often driven by a solar erupted event
such as a solar flare, EUV wave, or coronal jet or rain
\citep[][]{Zimovets15,Shen17,Reeves20,Zhang22}.

The kink oscillation observed here is triggered by a major flare and
it appears to propagate along several coronal loops.
Figure~\ref{corr} demonstrates the presence of phase difference in
the kink wave between two coronal loops, such as L2 and L1,
confirming that the kink oscillation propagates along different
coronal loops, namely, from L3 through L2 to L1. While
Figure~\ref{slit3} demonstrates that the initial short kink pulse
launched by the major flare is indeed a traveling wave in one
coronal loop. The traveling kink pulse is then evolved to a standing
kink oscillation in the coronal loop. To the best of our knowledge,
we observe the traveling kink oscillation evolving to the standing
kink wave for the first time.

We investigated the standing kink oscillation in coronal loop L2 in
detail because the oscillating loop has an entire profile with
a loop length of $\sim$221.3~Mm by assuming a
semi-circular loop shape (Figure~\ref{imag}). The oscillation
period is measured to be about 6.3~minutes, which is consistent with
previous findings in the period range of several minutes
\citep[e.g.,][]{Nakariakov99,Anfinogentov15,Su18,Nechaeva19,Ning22}.
The decaying time is estimated to 9.6$-$10.6~minutes and, thus, a
ratio of $\sim$1.5$-$1.7 was found between the decaying time and
oscillation period, similar to the average ratio found by
\cite{Nechaeva19}. Both the initial displacement and velocity
amplitude vary along the oscillating loop and they decrease when
the oscillating slits are far away from the major flare. For
instance, the displacement amplitude becomes from about 12.5~Mm at
slit S1 to around 6.1~Mm at slit S3 (Table~\ref{loop}). We cannot
find any signatures of phase difference among oscillating slits (S1,
S2, and S3) at multi-wavelength channels, suggesting that the kink
oscillation is indeed a fundamental mode. Based on the standing
kink-mode wave, a seismological inference of the magnetic field is
performed for the oscillating loop L2. The magnetic field strength
inside the coronal loop is estimated to 19.9$-$21.3~G, which is
consistent with previous estimations in coronal loops using
MHD coronal seismology
\citep{Nakariakov01,Aschwanden02,Yang20,Li23}. We want to stress
that MHD coronal seismology was not performed for oscillating loops
L1 and L3 because their loop profiles are incomplete, thus
their loop lengths cannot be measured.

The standing kink oscillation of coronal loop L3 appears to contain
multiple harmonics, since its displacement profile is very different
from a sine function. Therefore, we performed a wavelet transform
for the time series of the oscillating loop. Two dominant periods of
5.1~minutes (P$_1$) and 2.0~minutes (P$_3$) are identified in the
wavelet spectra and their period ratio (P$_1$/3P$_3$) is estimated
to 0.85, which agrees with previous findings in the decaying kink
oscillation \citep[cf.][]{Duckenfield19}. The departure of period
ratio from unity could be attributed to a density stratification of
the oscillating loop. Our observation implies that the kink wave
contains the fundamental and third harmonics. On the other hand, the
standing kink oscillation in coronal loop L1 also reveals a strong
signature of multiple harmonics, namely, non-sinusoidal displacement
profile. It is first in the fundamental mode along slit S1, and then
it contains multiple harmonics along slit S2. However, we could not
see an oscillation signature at slit S3, because loop L1 disappears
at slit S3. That is to say, it is evolved from the traveling kink
wave. Finally, we want to stress that the oscillation period of the
fundamental-mode kink wave becomes shorter and shorter from
oscillating loops L1 to L3, which could be attributed to the
observational fact that the oscillation period of kink waves is
strongly dependent on the loop length
\citep[e.g.,][]{Anfinogentov15,Goddard16,Li23}.

\section{Summary}
Based on the observation measured by SUTRI, SDO/AIA, GOES, and STIX,
we explored the decaying kink oscillation in three coronal loops. Our
main results are summarized as follows:

\begin{enumerate}

\item We first observe the evolution of a traveling kink pulse to a
standing kink wave in the coronal loop that has been triggered by a
major flare.

\item Based on the kink-mode wave, the Alfv\'{e}n speed and magnetic field
strength inside the oscillating loop were estimated to be about
950~km~s$^{-1}$ and 20~G, respectively.

\item The fundamental and third harmonics of kink wave were simultaneously
detected in the oscillating loop.

\end{enumerate}

\begin{acknowledgements}
We would like to thank the referee for his/her valuable
comments. This study is funded by the National Key
R\&D Program of China 2021YFA1600502 (2021YFA1600500), NSFC under
grant 11973092, 12073081, 11825301, 12273059, the Youth Fund of
Jiangsu No. BK20211402, the Strategic Priority Research Program on
Space Science, CAS, Grant No. XDA15052200 and XDA15320301. D.~Li is
also supported by Yunnan Key Laboratory of Solar Physics and Space
Science under the number YNSPCC202207. SUTRI is a collaborative
project conducted by the National Astronomical Observatories of CAS,
Peking University, Tongji University, Xi'an Institute of Optics and
Precision Mechanics of CAS and the Innovation Academy for
Microsatellites of CAS. SDO is NASA's first mission in the Living
with a Star program and AIA is an instrument onboard SDO. The STIX
instrument is an international collaboration between Switzerland,
Poland, France, Czech Republic, Germany, Austria, Ireland, and
Italy.
\end{acknowledgements}

\end{document}